\documentstyle[12pt]{article}

\def\A{{\cal A}}

\def\ag{{\cal A}/{\cal G}}
\def\agb{\overline{\ag}}

\def\c{{\cal C}}
\def\Cb{\overline{\cal C}}
\def\G{{\cal G}}

\def\HA{{\cal HA}}
\def\HAbar{\overline{\cal HA}}
\def\HG{{\cal HG}}
\def\L{{\cal L}}
\def\Lx{{\L}_{\xz}}
\def\X{{\cal X}}
\def\Xb{{\overline \X}}
\def\Xpb{{\overline{\X'}}}

\def\a{\alpha}
\def\b{\beta}

\def\S{M}

\def\w{\omega}

\def\0{\emptyset}
\def\Fun{{\rm Fun}}
\def\g{S}
\def\s{{S}}
\def\Tr{{\rm Tr}}
\def\C*{$C^{\star}$}

\def\Gb{{\overline \G}}
\def\Hom{{\rm Hom}}
\def\:{\ :}
\def\xz{x_0}
\def\Comp{{\bf C}}
\def\={\ =\ }
\def\qed{}% {${}\qquad\spadesuit$}
\def\Cyl{{\rm Cyl}}
\def\Cylb{{\overline \Cyl}}
\def\Fun{{\rm Fun}}

\newtheorem{theorem}{Theorem}
	\newtheorem{proposition}{Proposition}
	\newtheorem{lemma}{Lemma}

\begin{document}

\title{Projective techniques and functional integration
for gauge theories}
\author{Abhay Ashtekar${}^{1,2}$ and
Jerzy Lewandowski${}^{3,4}$}
\maketitle
\centerline{${}^{1}$Center for Gravitational Physics
and Geometry,}
\centerline{Physics Department, Penn State, University Park,
PA 16802-6300.}

\centerline{${}^{2}$Isaac Newton Institute for
Mathematical Sciences,}
\centerline{University of Cambridge, Cambridge CB3 OEH.}

\centerline{${}^{3}$Institute of Theoretical Physics,}
\centerline{Warsaw University, ul Hoza 69, 00-681 Warsaw.}

\centerline{${}^{4}$Erwin Schr\"odinger Institute for
Mathematical Physics}
\centerline{Pasteurgasse 6/7, A-1090 Vienna.}
\bigskip

{\Large{\bf Abstract}}
\bigskip

A general framework for integration over certain infinite dimensional
spaces is first developed using projective limits of a projective
family of compact Hausdorff spaces. The procedure is then applied to
gauge theories to carry out integration over the non-linear, infinite
dimensional spaces of connections modulo gauge transformations.  This
method of evaluating functional integrals can be used either in the
Euclidean path integral approach or the Lorentzian canonical approach.
A number of measures discussed are diffeomorphism invariant and
therefore of interest to (the connection dynamics version of) quantum
general relativity. The account is pedagogical; in particular prior
knowledge of projective techniques is not assumed.
\footnote{For the special JMP issue on {\it Functional Integration},
edited by C.  DeWitt-Morette.}

\goodbreak

\section{Introduction}

Theories of connections are playing an increasingly important role in
the current description of all fundamental interactions of Nature.
They are also of interest from a purely mathematical viewpoint. In
particular, many of the recent advances in the understanding of
topology of low dimensional manifolds have come from these theories.
To quantize such theories non-perturbatively, it is desirable to
maintain manifest gauge invariance. In practice, this entails working
directly on the quotient of the space of connections modulo (local)
gauge transformations. In particular, one has to develop an
integration theory on the quotient. Now, this is a {\it non-linear}
space with a rather complicated topology, while the standard
functional integration techniques \cite{K} are geared to linear
spaces. Therefore, a non-linear extension of these techniques is
needed. This task has been carried out in a series of papers over the
past two years [2-6].  (For earlier work with the same philosophy, see
\cite{AM,KK}.) The purpose of this article is to present a new,
self-contained treatment of these results.

Let us begin with a chronological summary of these developments. We
will also point out some of the limitations of the original methods
which will be overcome in the present treatment.

Fix an n-dimensional manifold $\S$ and consider the space $\A$ of
smooth connections on a given principal bundle $P(\S, G)$ over $\S$.
Following the standard terminology, we will refer to $G$ as the
structure group and denote the space of smooth maps from $\S$ to $G$
by $\G$. This $\G$ is the group of {\it local} gauge transformations.
If $\S$ happens to be a Cauchy surface in a Lorentzian space-time, the
quotient $\ag$ serves as the physical configuration space of the
classical gauge theory.  Following the familiar procedures from
non-relativistic quantum mechanics, one might therefore expect that
quantum states would be represented by square-integrable functions on
this $\ag$. In the sum over histories approach, $\S$ is taken to be
the Euclidean space-time and the quotient $\ag$ then represents the
space of physically distinct classical histories.  Therefore, one
might expect that the problem of constructing Euclidean quantum field
theory would reduce to that of finding suitable measures on this
$\ag$.

While these expectations point us in the right direction, they are not
quite accurate. For, it is well known that, due to the presence of an
infinite number of degrees of freedom, in quantum field theory one
must allow configurations and histories which are significantly more
general than those that feature in the classical field theories. (See,
e.g., \cite{GJ}.) For example, in scalar field theories in Minkowski
space, quantum states arise as functionals on the space of {\it
tempered distributions} on a $t= {\rm const}$ surface in space-time.
Similarly, in the Euclidean approach, measures are concentrated on
{\it distributional} histories; in physically interesting theories,
the space of smooth histories is typically a set of measure zero.  One
would expect the situation to be similar in gauge theories.
Unfortunately, since the spaces $\ag$ are non-linear, with complicated
topology, a canonical mathematical setting for discussing
``generalized connections modulo gauge transformations'' is not
available. For example, the naive approach of substituting the smooth
connections and gauge transformations in $\ag$ by distributional ones
does not work because the space of distributional connections does not
support the action of distributional local gauge transformations.

Recently, one such setting was introduced \cite{AI} using the basic
representation theory of \C*-algebras. The ideas underlying this
approach can be summarized as follows. One first considers the space
$\HA$ of functions on $\ag$ obtained by taking finite complex linear
combinations of finite products of Wilson loop functions $W_{\a}(A)$
around closed loops $\a$.  (Recall that the Wilson loop functions are
traces of holonomies of connections around closed loops; $W_{\a}(A) =
{\rm Tr}\ {\cal P}\ \oint_{\a} A dl$. Since they are gauge invariant,
they project down unambiguously to $\ag$.) By its very construction,
$\HA$ has the structure of a $\star$-algebra, where the involution
operation, $\star$, is just the complex-conjugation.  Since $G$ is
compact, $W_\alpha$ are bounded. This enables one to introduce an
obvious (i.e., the sup-) norm on $\HA$ and complete it to obtain a
\C*-algebra which we will denote by $\HAbar$. In the canonical
approach, $\HAbar$ is the algebra of configuration observables. Hence,
the first step in the construction of the Hilbert space of physical
states is the selection of an appropriate representation of $\HAbar$.
It turns out that every cyclic representation of $\HAbar$ by operators
on a Hilbert space is of a specific type \cite{AI}. The Hilbert space
is simply $L^2(\agb, d\mu)$ for some regular, Borel measure $\mu$ on a
certain completion $\agb$ of $\ag$ and, as one might expect of
configuration operators, the Wilson loop operators act just by
multiplication. Therefore, the space $\agb$ is a candidate for the
extension of the classical configuration space needed in the quantum
theory.

$\agb$ is called the Gel'fand spectrum of the \C*-algebra $\HAbar$ and
can be constructed purely algebraically: Its elements are the maximal
ideals of $\HAbar$.  As we just saw above, if $\S$ is a Cauchy
slice in a Lorentzian space-time, $\agb$ serves as the domain space of
quantum states. In the Euclidean approach, where $\ag$ is constructed
from connections on the Euclidean space-time, $\agb$ is the space of
generalized histories, over which one must integrate to calculate the
Schwinger functions of the theory. As in the linear theories, the
space $\ag$ of classical configurations/histories is densely embedded
in space $\agb$ of quantum configurations/histories. A key difference
from the linear case, however, is that, unlike the space of tempered
distributions, which is linear, $\agb$ is compact.

A central issue in this approach to quantum gauge theories is that of
obtaining a convenient characterization of $\agb$. For the cases when
the structure group $G$ is either $SU(n)$ or $U(n)$, this problem was
solved in \cite{AL1}. The resulting characterization is purely
algebraic.  Using piecewise analytic loops passing through an
arbitrarily chosen but fixed base point, one first constructs a group,
$\HG$, called the {\it hoop group}. One then considers the space ${\rm
Hom}(\HG, G)$ from $\HG$ to the structure group $G$. Every smooth
connection $A$ defines such a homomorphism via the holonomy map,
evaluated at the base point. However, ${\rm Hom}(\HG, G)$ has many
other elements. The space $\agb$ is then shown to be naturally
isomorphic to the space ${\rm Hom}(\HG, G)/ {\rm Ad G}$, where the
quotient by the adjoint action by $G$ just serves to remove the gauge
freedom at the base point. The proof of this characterization is
however rather long and relies, in particular, on certain results due
to Giles \cite{G} which do not appear to admit obvious extensions to
general compact gauge groups.

The space $\agb$ is very large. In particular, every connection on
{\it every} $G$-bundle over $\S$ defines a point in $\agb$. (In
particular, $\agb$ is independent of the initial choice of the
principal bundle $P(\S, G)$ made in the construction of the
holonomy algebra $\HA$.)  Furthermore, there are points which do not
correspond to {\it any} smooth connection; these are the generalized
connections (defined on generalized principal $G$-bundles \cite{Le})
which are relevant to only the quantum theory. There is a precise
sense in which this space provides a ``universal home'' for all
interesting measures \cite{ALMMT1}. In specific theories, such as
Yang-Mills, the support of the relevant measures is likely to be
significantly smaller. For diffeomorphism invariant theories such as
general relativity, on the other hand, the whole space appears to be
relevant.

Even though $\agb$ is so large, it is a compact, Hausdorff space.
Therefore, we can use the standard results from measure theory
directly. These imply that there is an interesting interplay between
loops and generalized connections. More precisely, every measure $\mu$
on $\agb$ defines a function $\Gamma_\mu$ on the space of multi-loops:
$$\Gamma_\mu (\a_1, ...,\a_n) := \int_{\agb}\
W_{\a_1}(\bar{A}) ... W_{\a_n}(\bar{A})\ d\mu \  ,\eqno(1.1)$$
where $W_{\a}(\bar{A})$ is the natural extension of $W_\a (A)$ to
$\agb$ provided by the Gel'fand representation theory. Thus, in the
terminology used in the physics literature, $\Gamma_\mu (\a_1, ...,
\a_n)$ is just the ``vacuum expectation value functional'' for the
Wilson loop operators. In the mathematics terminology, it is the
Fourier transform of the measure $\mu$. As in linear theories, the
Fourier transform $\Gamma_\mu$ determines the measure $\mu$
completely. Furthermore, {\it every} multi-loop functional $\Gamma$
which is consistent with all the identities in the algebra $\HA$ and
is positive defines a measure with $\Gamma$ as its Fourier transform.
Finally, the measure $\mu$ is diffeomorphism invariant if
$\Gamma_\mu$ is diffeomorphism invariant, i.e., is a functional of
(generalized) knots and links (generalized, because the loops $\a_1,
... \a_n$ are allowed to have kinks, intersections and overlaps.)

Several interesting regular Borel measures have been constructed on
$\agb$. The basic idea \cite{AL1} behind these constructions is to use
appropriate families of cylindrical functions, introduce cylindrical
measures and then show that they are in fact regular Borel measures.
Not surprisingly, the first non-trivial measure $\mu_o'$ to be
introduced on $\agb$ is also the most natural one: it is constructed
solely from the Haar measure on the structure group $G$.  Furthermore,
$\mu_o'$ has two attractive features; it is faithful and invariant
under the induced action of the diffeomorphism group of $\S$
\cite{AL1}. Marolf and Mour\~ao have analyzed properties of this
measure using projective techniques \cite{MM}.  They showed that $\ag$
is contained in a set of zero $\mu_o'$-measure; the measure is
concentrated on genuinely generalized connections. Using the relation
between the measures and knot invariants mentioned above, given a
suitable knot invariant, one can use $\mu_o'$ to generate other
diffeomorphism invariant measures \cite{ALMMT2}. The resulting family
of measures is very large. The next family is obtained by examining
the possible intersections and kinks in the loops and dividing
corresponding vertices into diffeomorphism invariant classes of
``vertex types.''  Then, by assigning to every $n$-valent vertex type
a measure on $G^n$, $G$ being the structure group, one can introduce a
measure on $\agb$.  This large class of measures was provided by Baez
\cite{Ba1,Ba2}. Another family arises from the use of heat kernel
methods on Lie groups \cite{AL2} and may be useful in field theories
on space-times with a fixed metric.  Finally, $\agb$ is known to admit
measures which are appropriate for the $SU(n)$ Euclidean Yang-Mills
theory in 2-dimensions, corresponding to space-times $\S$ with
topologies $R^2$ and $S^1\times R$ \cite{ALMMT1}. For loops without
self-intersections, the value of the generating functional
$\Gamma_\mu$ is given just by the exponential of (a negative constant
times) the area enlosed by the loop; confinement is thus manifest.
These measures are invariant under the induced action of
area-preserving diffeomorphisms of $\S$.

Thus, large classes of diffeomorphism invariant measures have been
constructed on $\agb$. It was somewhat surprising at first that $\agb$
admits {\it any} diffeomorphism invariant measures at all. Indeed,
there was a general belief that, just as translation-invariant
measures do not exist on infinite-dimensional topological vector
spaces, diffeomorphism invaraint measures would not exist on spaces of
connections modulo gauge transformations. That they exist is quite
fortunate because there are strong indications they will play an
important role in non-perturbative quantum general relativity
\cite{AA1}. Note finally that for each of these measures, the
computation of the expectation values of the Wilson loop operators
reduces to an integration over $k$-copies of the structure group, for
some $k$, and can therefore be carried out explicitly.

This concludes the general summary of recent developments. In this
paper, we will obtain these results from another perspective, that of
projective limits. The general idea of using these limits is not new:
it has been used already in \cite{MM} to analyze the support of the
measure $\mu_o'$ mentioned above. However, whereas the discussion of
\cite{MM} drew on earlier results \cite{AI,AL1} rather heavily,
here we will begin afresh. Thus, our treatment will serve three
purposes. First, it will enable us to provide a concise treatment of
the results discussed above without having to rely on external input,
such as the results of Giles \cite{G}, whose proofs are somewhat
involved.  Second, it will enable us to generalize the previous
results; we will be able to deal with all compact, connected gauge
groups at once.  Finally, this approach will enable us to show that
functional integration over non-linear spaces such as $\agb$ can be
carried out using methods that are closely related to those used in
the case of linear topological spaces \cite{K,DM1,DM2}, i.e., that
there is an underlying coherence and unity to the subject.

In section 2, we present a general framework in which the standard
projective techniques are applied to a projective family of compact
Hausdorff spaces. This framework is then used in section 3 to obtain
the space $\agb$ and to develop integration theory on it.

%\end{section}

\section{Projective Techniques: The general framework}

A general setting for functional integration over an infinite
dimensional, locally convex, topological space $V$ is provided by the
notion of ``projective families'' \cite{K,DM1,DM2} which are
constructed from the quotients, $V/\tilde{\g}$, of $V$ by certain
subspaces $\tilde{\g}$ (see section 2.1). We wish to extend this
framework gauge theories where the relevant space $\agb$ is {\it
non-linear}.  One approach to this problem is to modify the projective
family appropriately.  In section 2.1, we will begin by presenting
such a family; it will now be constructed using compact, Hausdorff
topological spaces.  In section 2.2, we will use some elementary
results from the Gel'fand-Naimark representation theory to unravel the
structure of the projective limit of this family. Section 2.3 provides
a general characterization of measures on the limiting space in terms
of measures on the projective family. Finally, in section 2.4, we
consider the action of a compact group on the family to obtain a
quotient projective family and show that the limit of the quotient
family is the same as the quotient of the limit of the original
family.  In application to gauge theory, the group action will be
given by the adjoint action of the structure group on holonomies,
evaluated at a fixed base point.)

While the primary application of the framework --discussed in the next
section-- is to gauge theories, the results are quite general and may
well be useful in other contexts.

\subsection{Projective family and the associated \C* algebra of
cylindrical functions}

Let $L$ be a partially ordered, directed set; i.e. a set equipped
with a relation `$\ge$' such that, for all $\g, \g'$ and $\g''$ in $L$
we have:
$$ \g \ge \g\ ;\quad
\g \ge \g'\ {\rm and}\  \g'\ge \g \Rightarrow \g =\g'\ ; \quad
\g\ge\g'\ {\rm and}\ \ \g'\ge\g''\ \Rightarrow \g\ge\g''\ ;\eqno(2.1a)$$
and, given any $\g', \g'' \in L$, there exists $\g \in L$ such that
$$ \g \ge \g' \quad {\rm and} \quad \g \ge \g''\ . \eqno(2.1b)$$
$L$ will serve as the label set.  A {\it projective family}
$(\X_\g,p_{\g\g'})_{\g,\g'\in L}$ consists of sets $\X_\g$ indexed by
elements of $L$, together with a family of surjective {\it projections},
$$p_{\g\g'}\:\ \X_{\g'}\rightarrow \X_{\g},\eqno(2.2)$$
assigned uniquely to pairs $(\g',\g)$ whenever $\g'\ge\g$
such that
$$p_{\g\g'}\circ p_{\g'\g''} = p_{\g\g''}.\eqno(2.3)$$

A familiar example of a projective family is the following. Fix a
locally convex, topological vector space $V$. Let the label set $L$
consist of finite dimensional subspaces $\g$ of $V^\star$, the
topological dual of $V$. This is obviously a partially ordered and
directed set. Every $\g$ defines a unique sub-space $\tilde{\g}$ of
$V$ via: $\tilde{v} \in \tilde{\g}\ \ {\rm iff}\ \ <v, \tilde{v}> =0\
\ \forall v\in \g$. The projective family can now be constructed by
setting $\X_\g = V/\tilde\g$. Each $\X_\g$ is a finite dimensional
vector space and, for $\g'\ge \g$, $p_{\g \g'}$ are the obvious
projections. As noted before, integration theory over infinite
dimensional topological spaces can be developed starting from this
projective family \cite{K,DM1,DM2}.

In this paper, we wish to consider another --and, in a sense,
complementary-- projective family which will be useful in certain
kinematically non-linear contexts. We will assume that $\X_\g$ are all
topological, compact, Hausdorff spaces and that the projections
$p_{\g\g'}$ are continuous.  We will say that the resulting pairs
$(\X_\g, p_{\g \g'})_{\g,\g' \in L}$ constitute a {\it compact
Hausdorff projective family}.  In the application of this framework to
gauge theories, carried out in the next section, the labels $S$ can be
thought of as general lattices (which are not necessarily rectangular)
and the members $\X_{\g}$ of the projective family, as the spaces of
configurations/histories associated with these lattices.  The
continuum theory will be recovered in the limit as one considers
lattices with increasing number of loops of arbitrary complexity.

Note that, in the projective family there will, in general, be no set
${\Xb}$ which can be regarded as the largest, from which we can
project to any of the $\X_\g$. However, such a set does emerge in an
appropriate limit, which we now define. The {\it projective limit}
$\Xb$ of a projective family $(\X_\g, p_{\g\g'})_{\g\g'\in L}$ is the
subset of the Cartesian product $\times_{\g\in L}\X_\g$ that satisfies
certain consistency conditions:
$$\Xb\ :=\ \{(x_\g)_{\g\in L}\in
\times_{\g\in L}\X_\g\ :\ \g'\ge \g \Rightarrow p_{\g\g'}x_{\g'} =
x_\g\}.  \eqno(2.4) $$
(This is the limit that will give us the continuum gauge theory in the
next section.)  We provide $\Xb$ with the product topology that
descends from $\times_{\g\in L}\X_\g$. This is the {\it Tychonov
topology}.  The topology of the product space is known to be compact
and Hausdorff.  Furthermore, as noted in \cite{MM}, $\Xb$ is closed in
$\times_{\g\in L} \X_\g$, whence $\Xb$ is also compact (and
Hausdorff). We shall establish this property of $\Xb$ independently in
section 2.2. For now, we only note that the limit $\Xb$ is naturally
equipped with a family of projections:
$$ p_{\g}\ : \ \Xb \rightarrow \X_{\g}, \ \ p_{\g} ((x_{\g'})_{\g'
\in L}):= x_{\g}\ .\eqno(2.5)$$

Next, we introduce certain function spaces. For each $\g$ consider
the space $C^0(\X_\g)$ of the complex valued, continuous functions on
$\X_\g$. In the union
$$\bigcup_{\g\in L} C^0(\X_{\g})$$
let us define the following equivalence relation. Given $f_{\g_i}\in
C^0(\X_{\g_i})$, $i=1,2$, we will say:
 $$f_{\g_1} \ \sim\ f_{\g_2}\ \ \ {\rm if}\ \
\ p_{\g_1\g_3}^* \ f_{\g_1}\ =\ p_{\g_2\g_3}^*\ f_{\g_2}\eqno(2.6)$$
for every $\g_3\ \ge \g_1,\g_2$, where $p^*_{\g_1,\g_3}$ denotes the
pull-back map from the space of functions on $\X_{\g_1}$ to the space
of functions on $\X_{\g_3}$. Note that to be equivalent, it is
sufficient if the equality (2.6) holds {\it just for one} $\g_3\ \ge
\g_1,\g_2$. To see this, suppose that (2.6) holds for $\g_1, \g_2$
and $\g_3$ and let ${\g_4}\ge \g_1,\g_2$. Take any $\g_5\ge\g_i$,
$i=1,2,3,4$.  Then
$$p_{\g_4\g_5}^*p_{\g_1\g_4}^*f_{\g_1} =
p_{\g_3\g_5}^* p_{\g_1\g_3}^*f_{\g_1} =
p_{\g_3\g_5}^*p_{\g_2\g_3}^*f_{\g_2} =
p_{\g_4\g_5}^*p_{\g_2\g_4}^*f_{\g_2}.\eqno(2.7)$$
Since $p_{\g\g'}^*:C^0(\X_\g)\rightarrow C^0(\X_{\g'})$ is an
embedding, (2.7) implies
$$ p_{\g_1\g_4}^*f_{\g_1} = p_{\g_2\g_4}^*f_{\g_2}.\eqno(2.8)$$

Using the equivalence relation we can now introduce the set of
{\it cylindrical functions} associated with the projective
family $(\X_\g,p_{\g\g'})_{\g,\g'\in L}$,
$$\Cyl(\Xb) \ := \ \big( \bigcup_{\g\in L}C^0(\X_\g)\ \big)
\ /\ \sim.\eqno(2.9)$$
The quotient just gets rid of a redundancy: pull-backs of functions
from a smaller set to a larger set are now identified with the
functions on the smaller set.  Note that in spite of the notation, as
defined, an element of $\Cyl(\Xb)$ is {\it not} a function on ${\Xb}$;
it is simply an equivalence class of continuous functions on some of
the members $\X_\g$ of the projective family. The notation is,
however, justified because, as we will show in the section 2.2, one
{\it can} identify elements of $\Cyl(\Xb)$ with continuous functions
on $\Xb$.

Henceforth we shall denote the element of $\Cyl(\Xb)$ defined by
$f_\g\in C^0(\X_\g)$ by $[f_\g]_\sim$. We have:

\begin{lemma}{\rm :} Given any $f,g\in \Cyl(\Xb)$, there exists
$\g\in L$ and $f_\g, g_\g\in C^0(\X_\g)$ such that
$$ f = [f_\g]_\sim,\ \ \ g = [g_\g]_\sim\eqno(2.10)$$
\end{lemma}

{\it Proof} : Choose any two representatives $f_{\g_1}\in f$ and
$g_{\g_1}\in g$; there exists $\g\in L$ such that $\g\ge\g_i$,
$i=1,2$. Take  $f_\g = p_{\g_1\g}^*f_{\g_1}$ and
$g_\g = p_{\g_2\g}^*g_{\g_2}$.

\qed
\medskip

We will conclude this sub-section with a proposition that collects
elementary properties of the cylindrical functions which makes it
possible to construct a \C* algebra out of $\Cyl(\Xb)$.

\begin{proposition}{\rm :}

\noindent (i) Let  $f,g\in\Cyl(\Xb)$; then the following
operations are well defined:
$$ f+g\ :=\ [f_\g + g_\g]_\sim,\ \ \ fg\ :=\ [f_\g g_\g]_\sim,
\eqno(2.11a)$$
$$af\ := \ [af_\g],\ \ \ f^\star \ := [f_\g^\star]\eqno(2.11b)$$
where $\g$ is any element of $L$ given by Lemma 1,
$a\in \Comp$ and $\star$ within the bracket is the complex
conjugation.

\noindent (ii) A constant function belongs to $\Cyl(\Xb)$.

\noindent (iii) Let $f_\g\in C^0(\X_\g)$, $f_{\g'}\in
C^0(\X_{\g'})$ and $f_\g\sim f_{\g'}$; then
$$\sup_{x_\g\in \X_\g}|f_\g(x_\g)|\ =\
\sup_{x_{\g'}\in \X_{\g'}}|f_{\g'}(x_{\g'})|.\eqno(2.11c)$$
\end{proposition}
The proofs are  consequence of  Lemma 1 and use an argument
similar to that proving (2.8).
\medskip

It follows from Proposition 1 that the set of cylindrical functions
$\Cyl(\Xb)$ has the structure of a $\star$--algebra with respect to
the operations defined in (i). The algebra contains the identity and
we define on it a norm $\|\ \|$ to be
$$\|[f_\g]_\sim\|\ := \sup_{x_\g\in\X_\g}|f_\g(x_\g)|.
\eqno(2.12)$$
The norm is well defined according to part (iii) of Proposition 1.
Consider the Banach algebra $\Cylb(\Xb)$ obtained by taking the Cauchy
completion of $\Cyl(\Xb)$. Since
$$\|f^\star\|\ =\ \|f\|,\eqno(2.13)$$
$\Cylb(\Xb)$ is a \C*-algebra. By construction, it is Abelian and has
an identity element.  We shall refer to it as the {\it \C*-algebra
associated with the projective family} $(\X_\g,p_{\g\g'})_{\g\g'\in\G}$.

%\end{subsection}
\subsection{The Gel'fand spectrum of $\Cylb(\Xb)$.}

A basic result in the Gel'fand-Naimark representation theory assures
us that every Abelian \C*-algebra $\Cb$ with identity is realized as
the \C*-algebra of continuous functions on a compact Hausdorff space,
called the {\it spectrum} of $\Cb$. Furthermore, the spectrum can be
constructed purely algebraically: Its elements are the maximal ideals
of $\Cb$, or, equivalently, the $\star$-preserving homomorphisms from
$\Cb$ to the $\star$-algebra of complex numbers.  In this sub-section,
we will use this result to show that the spectrum of $\Cylb(\Xb)$ is
precisely the projective limit $\Xb$ of our projective family $(\X_\g,
p_{\g \g'})_{\g,\g'\in L}$. This will establish that, although the
\C*-algebra $\Cylb(\Xb)$ was constructed using the projective family,
its elements can be identified with continuous functions on the
projective limit. In section 3, this result will enable us to obtain a
characterization of the space $\agb$ in a way that is significantly
simpler and more direct than the original procedure \cite{AI,AL1}.

To show that $\Xb$ is the spectrum of $\Cyl(\Xb)$, it is sufficient to
establish the appropriate isomorphism from $\Cyl(\Xb)$ to the
\C*-algebra $C^0(\Xb)$ of continuous functions on $\Xb$. There is an
obvious candidate for this isomorphism. To see this, recall first,
from Eq (2.5), that there is a natural projection map $p_{\g}$
from $\Xb$ to $\X_{\g}$. It is easy to check that
$$f_{\g_1}\sim f_{\g_2}\ \ \ \Rightarrow\ \ \ p_{\g_1}^*f_{\g_1}\=\
p_{\g_2}^*f_{\g_2}\ .\eqno(2.14)$$
Hence, we have a representation of the algebra of cylindrical
functions $\Cyl(\Xb)$ by functions on the projective limit $\Xb$
$$\Cyl(\Xb)\ni f=[f_\g]_\sim\ \mapsto F(f)\:=\ p_\g^*f_\g\in
\Fun(\Xb),
\eqno(2.15)$$
Furthermore, this map satisfies $\sup_{x\in\Xb}|F(f)(x)|\ \le \
\|f\|$.  Hence it extends to the completion $\Cylb(\Xb)$. Thus
the map $F$ of Eq (2.15) is a candiate for the isomorphism we are
seeking.  Unfortunately, it is not obvious from the definition of the
map that is it surjective; i.e. that {\it every} continuous function
on $\Xb$ arises from the pull-back of a function on a $\X_{\g}$. We
will therefore follow a different route towards our goal, using
directly the algebraic definition of the spectrum. The final result
will, in particular, establish that the map $F$ has the desired
properties.

As mentioned above, the Gel'fand spectrum of $\Cylb(\Xb)$ is the set
of homomorphisms $\Hom (\Cylb(\Xb),\- \Comp)$ from the $\star$-algebra
$\Cylb(\Xb)$ into the $\star$-algebra of complex numbers. Now, since
$\Cyl(\Xb)$ is dense in $\Cylb(\Xb)$, every {\it continuous},
$\star$-operation preserving homomorphism from $\Cyl(\Xb)$ extends
uniquely to an element of $\Hom (\Cylb(\Xb),\- \Comp)$. Hence, to find
the spectrum of $\Cylb(\Xb)$ it suffices to find the space
$\Hom_o(\Cyl(\Xb), \Comp)$ of the continuous homomorphisms from the
$\star$-algebra $\Cyl(\Xb)$ to complexes. Our first observation is
that there is a natural map $\phi$ from $\Xb$ to $\Hom_o(\Cyl(\Xb),
\Comp)$,
$$\phi\ :\ \Xb\rightarrow \Hom_o(\Cyl(\Xb), \Comp),
\ \ \ \phi(x)(f)\ :=\ f_\g(p_\g(x))\ ,  \eqno(2.16)$$
where $x\in\Xb$ and $\g$ is any element of $L$ such that $f =
[f_\g]_\sim$ with some $f_\g\in C^0(\X_\g)$, since by definition of
the projective limit, the right hand side of the Eq (2.16) is
independent of a choice of $\g$. We now show that $\phi$ is in fact a
homeomorphism:

\begin{theorem}{\rm :} Let $\Cyl(\Xb)$ be the $\star$--algebra of
cylindrical functions of a Hausdorff, compact projective family
$(\X_\g, p_{\g\g'})_{\g, \g'\in L}$; then the map $\phi$ of Eq (2.16)
is a homeomorphism from the projective limit $\Xb$ equipped with the
Tychonov topology onto the Gel'fand spectrum of $\Cylb(\Xb)$ with its
Gel'fand topology.
\end{theorem}

{\it Proof} : The map $\phi$ is an injection because, for each
$\g\in L$, $C^0(\X_\g)$ separates the points of $\X_\g$ .
To see that $\phi$ is a surjection we construct the inverse
mapping. Let us fix
$$h\in\Hom_o(\Cyl(\Xb), \Comp)\ . $$
For each $\X_\g$, $h$ defines $h_\g \in \Hom_o(C^0(\X_\g),
\Comp)$ by
$$h_\g(f_\g)\:=\ h([f_\g)]_\sim)\ .$$
Now, since $\X_\g$ is compact and Hausdorff, given a $h_\g$ there
exists $x^h_\g\in \X_\g$ such that
$$h_\g(f_\g)\ = \ f_\g(x^h_\g)\ \ \ {\rm for}\ \ {\rm every}
f_\g\in C^0(\X_\g). $$
In this way, for every $\g\in L$ we have assigned to $h$ a point
$x^h_\g$ lying in $\X_\g$. Pick any $\g_1,\g_2\in L$ such that
$\g_2\ge\g_1$ and find the corresponding points $x^h_{\g_i}$. Then,
for each $f_{\g_1}\in C^0(\X_{\g_1})$, we have
$$f_{\g_1}(p_{\g_1\g_2}(x^h_{\g_2}))\ =\
p_{\g_1\g_2}^*f_{\g_1}((x^h_{\g_2}))\ =\ h_{\g_2}(f_{\g_2})\ =\
h_{\g_1}(f_{\g_1})\ =\ f_{\g_1}(x^h_{\g_1}),\eqno(2.17)$$
from which we conclude that
$$p_{\g_1\g_2}(x^h_{\g_2})\ =\ x^h_{\g_1}.\eqno(2.18)$$
Hence, $(x_\g)_{\g\in L}\in\Xb$ and
$$h([f_\g]_\sim)\ =\ f_\g(x_\g).\eqno(2.19)$$
Thus, we have proven that $\phi$ is a bijection.

The Gel'fand transform is a representation of $\Cylb(\Xb)$,
$$\check{G} \:\ \Cylb(\Xb)\ \rightarrow\ \Fun(\Hom_o(\Cyl(\Xb),\Comp ),
\ \ \ \check{G}(f)(h)\:=\ h(f)\ .\eqno(2.20)$$
The Gel'fand topology on $\Hom_o(\Cyl(\Xb), \Comp )$ is the weakest
topology with respect to which all the functions
$\check{G}(\Cyl(\Xb))$ are continuous. On the other hand, expressed in
terms of functions, the Tychonov topology on $\Xb$ is generated by the
functions $\bigcup_{\g\in L}p_\g^*(C^0(\X_\g))$.  Because the map
$\phi^*$ is a bijection of $\check{G}(\Cyl(\Xb))$ onto $\bigcup_{\g\in
L}p_\g^*(C^0(\X_\g))$, $\phi^*$ carries the Gel'fand topology into
that of Tychonov and ${\phi^{-1}}^*$ maps the Tychonov topology into
that of Gel'fand.
\qed

\bigskip
Theorem 1 has several interesting implications. We will conclude this
sub-section by listing a few.
\begin{enumerate}
\item We can now conclude that the representation map $F$ from
$\Cyl({\Xb})$ to functions on $\Xb$ of Eq (2.15) is faithful.
\item Second, the map $F$ preserves the norms:
$$\sup_{x\in\Xb}|F(f)(x)|\ = \ \|f\|\ ; $$
$F$ can therefore be extended to an isomorphism between $\c(\Xb)$ and
$C^0 (\Xb)$.
\item Third, we can establish two properties of the projective
limit $\Xb$:\\
%\begin{proposition}{\rm :}
\smallskip %\noindent
(a) The topology of the pointwise convergence in $\Xb$ (i.e.,
$(x^n_\g)_{\g\in L}\ \rightarrow\ (x^0_\g)_{\g\in L}$ iff
$x^n_\g\ \rightarrow\ x^0_\g$ for every $\g\in L $) is Hausdorff
and compact.  This follows from the fact that the topology in question
is just the Tychonov topology, and, from theorem 1, $\Xb$ with Tychonov
topology is homeomorphic with the Gel'fand spectrum which is
compact, and Hausdorff.\\
\smallskip %\noindent
(b)the map  $\Xb\ni (x_{\g'})_{\g'\in L}
\mapsto x_\g\in\X_\g$ is surjective for every $\g\in L$.
%\end{proposition} {\it Proof.}
This statement follows from the property 2 of the map $F$. Indeed, let
$x^0_\g\in \X_\g$. There exists an everywhere positive $f_\g\in
C^0(\X_\g)$ such that $f_\g(x_\g)-f_\g(x^0_\g) < 0$, for every
$x_\g\in L$.  Hence $\sup_{x\in\Xb} |F(f)(x)| = f(x^0_\g)$. But
since $F(f)$ is a continuous function on a compact space, there exists
$x^1\in\Xb$ such that $F(f)(x^1)= \sup_{x\in\Xb}|F(f)(x)|$. Then
$f_\g(x^1_\g)= f_\g(x^0_\g)$ which implies that $x^0_\g = p_\g (x^1)$.
%\qed\smallskip
\end{enumerate}

\subsection{Regular Borel measures on the projective limit}

The projective limit, $\Xb$, is a compact, Hausdorff space in its
natural\\ --Tychonov-- topology. Hence we can apply the standard results
from measure theory to it. What we need is a procedure to construct
interesting regular Borel measures on $\Xb$. To simplify this task,
in this sub-section we will obtain a convenient characterization of
these measures.

Let us begin with a definition. Let us assign to each $\g \in L$, a
regular Borel, probability (i.e., normalized) measure, $\mu_\g$ on
$\X_{\g}$. We will say that this is a {\it consistent family of
measures} if
$$ (p_{\g\g'})_{*}\ \mu_{\g'}\ =\ \mu_{\g}\ . \eqno(2.21)$$
Using this notion, we can now characterize measures on $\Xb$:

\begin{theorem}{\rm :} Let $(\X_\g, p_{\g \g'})_{\g\g' \in
L}$ be a compact, Hausdorff projective family
and $\Xb$ be its projective limit.

\smallskip
\noindent (a) Suppose $\mu$ is a regular Borel, probability
measure on $\Xb$; then $\mu$ defines a consistent family of
regular, Borel, probability measures, given by:
$$\mu_\g\:=\ {p_\g}_*\mu;\eqno(2.22)$$

\smallskip
\noindent (b) Suppose $(\mu_\g)_{\g,\g'\in L}$ is a consistent family
of regular, Borel, probability measures. Then there is a unique
regular, Borel, probability measure $\mu$ on $\Xb$ such that
$({p_\g})_*\ \mu = \mu_\g$;

\smallskip
\noindent(c) $\mu$ is faithful if  $\mu_\g\ :=\
({p_\g})_*\ \mu$  is faithful for every $\g\in L$.
\end{theorem}

{\it Proof} :

\noindent(a) Fix  $\X_\g$ and define the following
functional on $C^0(\X_{\g})$:
$$C^0(\X_\g)\ \ni\ f_\g\ \mapsto\ \int_\Xb d\mu [f_\g]_\sim.
\eqno(2.23)$$
Being linear and positive, the functional defines a regular Borel
measure $\mu_\g$ on $\X_\g$ such that
$$\int_{\X_\g} d\mu_\g f_\g\ =\
\int_\Xb d\mu [f_\g]_\sim.\eqno(2.24)$$
The unit function on $\X_\g$, $I_\g(x_\g)\ =\ 1$
is lifted to $[I_\g]_\sim\in\Cyl(\Xb)$, $[I_\g]_\sim(x)\ =\ 1$.
Hence, since $\mu$ is a probability measure, so is $\mu_\g$.

\smallskip
\noindent(b) Given a self consistent family of measures,  we
will define, now on $C^0(\Xb)$, a functional which is positive and
linear. Using Theorem 1, let us identify the space of cylindrical
functions $\Cyl(\Xb)$ with a dense subspace of $C^0(\Xb)$.  Let
$f\in\Cyl(\Xb)\subset C^0(\Xb)$. Then $f=[f_\g]_\sim$ for some
$\g\in\Gamma$ and $f_\g\in C^0(\X_\g)$. It follows from the definition
of the equivalence relation $\sim$ and the consistency conditions
(2.21), that the following functional is well defined:
$$\Gamma(f)\ :=\ \int_{\X_\g}d\mu_\g f_\g\ .  \eqno(2.25)$$
Obviously, $\Gamma$ is linear and positive on $\Cyl(\Xb)$, hence
continuous. Because $\Cyl(\Xb)$ is dense in $C^0(\Xb)$ the functional
extends to the positive linear functional defined on the entire
$C^0(\Xb)$. Finally, there exists a regular Borel measure $\mu$ on
$\Xb$ such that
$$\int_\Xb d\mu f\ =\ \Gamma(f),\ \ \ f\in C^0(\Xb).$$
This concludes the proof of (b).

\noindent(c) Suppose  $\mu$ is faithful. If $f_\g\in C^0(\X_\g)$
then,
$$\int_{\X_\g} d\mu_\g f_\g^\star f_\g \ =\
 \int_\Xb d\mu ({p_\g}^*f_\g)^\star ({p_\g}^*f_\g).$$
Therefore the vanishing of the left hand side implies the
vanishing of ${p_\g}^*f_\g$. But since we have proved in
Theorem 1 that the map (2.15) is an injection, it follows that
$f_\g=0$.

Suppose finally that the measures $\mu_\g$ are all faithful. Then the
linear functional $\Gamma$ (2.25) is strictly positive on $\Cyl(\Xb)
\subset C^0(\Xb)$. Since $C^0(\Xb)$ is the closure of $\Cyl(\Xb)$,
it follows from general results on \C* that $\Gamma$ extends to a
strictly positive linear functional on $C^0(\Xb)$.  Alternatively, we
can establish this result using measure theory.  The product topology
on $\Xb$ is generated by the cylindrical open sets: pullbacks
$p_\g^*U_\g$ of the open subsets $U_\g$ in $\X_\g$, $\g\in\Gamma$. It
is easy to verify that the intersection of two open cylindrical sets
is a cylindrical set. Hence, every non-empty open set in $\Xb$
contains a non-empty cylinder set. Let $U$ be a non-empty open set in
$\Xb$ and let $U'\subset U$ be cylindrical and non-empty.  Then
$$\mu(U)\ \ge\ \mu(U')\ > \ 0$$
which concludes the proof of (c).

\qed\medskip\noindent
{\bf Remark}: Note that, in establishing Theorem 2, we have used the
results of section 2.1 in an essential way. First, for the generating
function $\Gamma$ of (2.25) to be well-defined, it is essential that the
map (2.15) be one to one, which in turn is a consequence of Theorem 1.
This property is also used in showing that the faithfulness of $\mu$
implies that of the projection ${p_\g}_*\mu$.

\subsection{Quotient of a projective family.}

Let now us suppose that a compact, Hausdorff topological group $G$
acts on each component $\X_\g$ of a projective family $(\X_\g,
p_{\g\g'})_{\g,\g'\in L}$,
$$\X_\g\times G\ \ni\ ( x_\g, g_\g)\ \mapsto\ x_\g\ g_\g \ \in\
\X_\g. \eqno(2.26)$$
Suppose further that the map (2.26) is continuous and the projections
$p_{\g\g'}$ are $G$-equivariant:
$$p_{\g\g'}(x_{\g'}\ g_{\g'})\ =\ x_\g\ g_{\g}  .\eqno(2.27)$$
Then, we will say that the $G$-action on the projective family is {\it
consistent}. In this case, the initial projective family descends to a
compact, Hausdorff projective family of the quotients, $(\X_\g/G,
p_{\g\g'})_{\g,\g' \in \Gamma}$.

While working simultanously with two projective families $(\X_\g,
p_{\g\g'})_{\g,\g'\in \Gamma}$ and $(\X_\g/G, p_{\g\g'})_{\g,\g'\in
\Gamma}$ we will mark the objects corresponding to the quotient family
with the subscript ${}_G$. Thus for example, the two spaces of
cylindrical functions will be denoted by $\Cyl$ and $\Cyl_G$,
respectively.  However, for simplicity of notation, we will use the
same symbol $p_{\g \g'}$ to denote the projection maps of both
families; the intended projection should be clear from the context.

Now, the equivariance property (2.27) implies that the action (2.26)
of $G$ on the projective family $(\X_\g, p_{\g \g'})_{\g,\g' \in
L}$ induces a natural action of $G$ on its projective limit $\Xb$
$$ \Xb\ni (x_\g)_{\g\in\Gamma}\mapsto (x_\g \ g_\g)_{\g\in\Gamma}
\in\Xb.\eqno(2.28)$$
Taking the quotient, we obtain a compact Hausdorff space $\Xb/G$. On
the other hand, we can also take the projective limit of the quotient
family $(\X_\g/G, p_{\g\g'})_{\g,\g'\in \Gamma}$, to obtain a compact,
Hausdorff space $\Xb_G$. It is therefore natural to ask for the relation
between the two. Not surprisingly, they turn out to be isomorphic. This
is the main result of this sub-section.

We begin by noting that there is a natural sequence of maps:
$$\Xb\ \rightarrow\ \Xb/G \rightarrow \Xb_G,\ \
\ (x_\g)_{\g\in\Gamma}\mapsto [(x_\g)_{\g\in\Gamma}] \mapsto
([x_\g])_{\g\in\Gamma},\eqno(2.29)$$
where the square bracket denotes the orbits of $G$ (in $\Xb$ and in
$\X_\g$ respectively). Using this sequence, we can now state  the
theorem:

\begin{theorem}{\rm :} Let $G$ and $(\X_\g, p_{\g\g'})_{\g,\g'\in
\Gamma}$ be a Hausdorff compact group and a compact projective
family respectively. Suppose $G$ acts consistently on $(\X_\g,\
p_{\g\g'})_{\g,\g'\in L}$; then, the map
$$
\phi:\ \ \Xb/G\ni [(x_\g)_{\g\in\Gamma}]\ \ %{\buildrel \phi \over
{\mapsto}\ \  ([x_\g])_{\g\in\Gamma}\in \Xb_G\eqno(2.30)$$
is a homeomorphism.
\end{theorem}

{\it Proof} : This follows from the following two lemmas.

\begin{lemma}{\rm :} The pullback map
$$\phi^*\: \Cylb_G(\X_G)\ \rightarrow\ C^0(\Xb/G)\eqno(2.31)$$
is a bijection.
\end{lemma}

{\it Proof} : Let $f_G=[ {f_G}_\g]_{\sim_G}\in \Cyl_G$ where ${ f_G}_\g
\in C^0(\X_\g/G)$. Then, $\phi^*f_G$ coincides with the projection to
$\Xb/G$ of the function $[f_\g]_\sim\in C^0(\Xb)$, $f_\g$ being the
$G$ invariant function on $\X_\g$ corresponding to the function
${f_G}_\g$. Hence $\phi^*$ carries $\Cyl_G$ (thought of as the set of
$G$ invariant elements of $\Cyl$) injectively into $C^0(\Xb/G)$.
Moreover, the image, $\phi^*\Cyl_G$ is dense in $C^0(\Xb/G)$. Finally,
$\phi^*$ preserves the norm: $\|f_G\|_G = \|[f_\g]_\sim\| =
\|\phi^*f_G\|$, where, the last term denotes the sup-norm in
$C^0(\Xb/G)$, and is therefore a continuous map. Hence the result.
\medskip
Next, we have:
\medskip

\begin{lemma}{\rm :} Let $W$ and $Z$ be two Hausdorff, compact
spaces. Suppose there exists a map $\phi:W\rightarrow Z$ is such that
$\phi^*:C^0(Z) \rightarrow C^0(W)$ is a bijection. Then $\phi$ is a
homeomorphism.
\end{lemma}
{\it Proof} : The map $\phi$ has to be injective because $W$ is
Hausdorff. Suppose that $z\in Z$ is not in the image of $\phi$.  $z$
nonetheless defines an element of the spectrum of $C^0(Z)$, i.e., a
($\star$-preserving) homomorphism from $C^0(Z)$ to the space of
complex numbers. Since $\phi^*$ is a bijection, here must exist
$z'\in\phi(W)$ which represents the same homomorphism. But this
contradicts the hypothesis that $Z$ is Hausdorff, so $\phi(W)=Z$.
Thus $\phi$ is also surjective. Finally, we establish the continuity
of $\phi$. We have just established that $\phi^{-1}$ exist. The
hypothesis of the lemma implies that ${\phi^{-1}}^*\:\ C^0(W)\
\rightarrow\ C^0(Z)$ is also bijective.  Since the topologies are given by
the space of all continuous functions, it is clear that $\phi$ is
a homeomorphism.
\smallskip

Combining the above lemmas we complete the proof of the
theorem.
\qed
\medskip

We will conclude with an observation. Define a {\it group projective
family} to be a projective family $(\G_\g, p_{\g\g'})_{\g,\g'\in L}$
such that every $\G_\g$ is a topological group and the projections
$p_{\g\g'}$ are homomorphisms. It is not hard to see that the
projective limit $\G$ provided with the natural topology is a
topological group.  Furthermore, we have:
\medskip
\begin{proposition}{\rm :} Suppose  $\Gb$ is the projective limit
of a group projective family $(\G_\g, p_{\g\g'})_{\g,\g' \in
L}$. If all the  components  $\G_\g$ are compact, then
$\Gb$ is a compact topological group.
\end{proposition}
Therefore, if to a projective family $(\X_\g, p_{\g\g'})_{\g,\g'\in
\Gamma}$ there is associated a group projective family $(\G_\g,
p_{\g\g'})_{\g,\g'\in L}$ of groups of motions of $\X_\g$s which act
consistently with the projections the projective limit of the
quotients is again the quotient of the projective limits.

\section{Application to gauge-theories}

We will now apply the results of section 2 to gauge theories.

Section 3.1, introduces the relevant projective family. We begin with
the notion of based loops and regard two as being equivalent if the
holonomies of any smooth connection around them are the same. Each
equivalence class is called a hoop. The space of equivalence classes
has the structure of a group, which is called the hoop group and
denoted by $\HG$. Subgroups $\g$ of $\HG$, generated by a finite
number of hoops, will serve as the labels for our family.  The set
$\X_\g$ will consist of all homomorphisms ${\rm Hom}(\g, G)$ from $\g$
to the structure group $G$. If we regard $\g$ as forming a
(``floating'', i.e., irregular) lattice, $\X_{\g}$ is in essence the
space of configurations in the canonical approach and of histories in
the Euclidean approach to the $\X_{\g}$-lattice gauge theory. The
projective limit $\Xb$ then serves as the corresponding space for the
continuum theory.  In section 3.2, we consider the \C*-algebra
$\Cyl(\Xb)$ of cylindrical functions associated with this projective
family. We show that this is naturally isomorphic with a certain
\C*-algebra of functions on $\ag$ which is known \cite{AL1} to be
naturally isomorphic to the holonomy algebra $\HAbar$ of Wilson loop
functions of the continuum gauge theory.  Theorem 1 now implies that
the projective limit $\Xb$ of our projective family is naturally
isomorphic with the Gel'fand spectrum $\agb$ of $\HAbar$. This
provides a complete characterization of the spectrum without external
inputs. Furthermore, since this characterization does not use results
--such as those of Giles \cite{G}-- which are tied to $SU(n)$ or
$U(n)$, we will now be able to treat all compact, connected structure
groups $G$ at once. Finally, this approach enables us to use directly
the projective techniques of section 2.3, thereby streamlining the
task of introducing measures on $\agb$. This construction is carried
out in section 3.3.

The specific projective family we use here was first introduced
explicitly by Marolf and Mour\~ao \cite{MM} using techniques developed
in \cite{AL1}. Our present treatment is a continuation of their work.
However, to establish certain properties of the projective limit $\Xb$
--such as the fact that $\Xb$ is actually larger than $\ag$-- they had
to refer to the results of \cite{AI,AL1} which in turn depend on the
results due to Giles \cite{G}.  Using the general framework developed
in section 2, we will now be able to establish these properties more
directly.

Another projective family, based on graphs in $\S$ rather
than subgroups of $\HG$, was introduced by Baez \cite{Ba1,Ba2}.
It is better suited for constructing Baez measures and for
introducing differential geometry on $\agb$ \cite{AL2}.
\goodbreak

\subsection{The projective family}
\medskip
\noindent{\bf The set of labels.}

Let us begin by recalling the notion of based loops. Let $\S$ be an
analytic manifold and consider continuous, piecewise analytic
($C^\w$), parametrized {\it curves}, i.e., maps
$$p\ :\ \  [0, s_1]\cup \dots\cup[s_{n-1}, 1]\rightarrow \S$$
which are  continuous on the whole domain and $C^\w$  on the closed
intervals $[s_k, s_{k+1}]$. Given two curves $p_1:\ [0, 1]
\rightarrow\ \S$ and  $p_2:\ [0, 1] \rightarrow\ \S$ such that $p_1(1)
=p_2(0)$, we will denote by $p_2\circ p_1$ the natural composition:
$$p_2\circ p_1(s) = \cases{p_1(2s),& for $s\in [0, {1\over 2}]$ \cr
       p_2(2s-1), &  for $s\in [{1\over 2}, 1]$.\cr} $$
The {\it inverse} of a curve $p\ \colon\ [0, 1] \rightarrow \S$ is a
curve given by
$$p^{-1}(s) := p(1 - s).$$
A curve which begins and ends at the same point is called a {\it
loop}.  Fix, once and for all, a point $\xz\in \S$. Denote by $\Lx$
the set of (continuous, piecewise $C^\w$) loops which are based at
$\xz$, i.e., which start and end at $\xz$.

Given a compact, connected Lie group $G$, two loops $\a_1, \a_2$ in
$\S$ will be said to be $G$-holonomy equivalent if
for every Lie algebra valued 1-form $A$ on $\S$, we have
$${\cal P}\exp\int_{\a_1} Adl\ =\ {\cal P}\exp\int_{\a_2} Adl\ , $$
where the holonomies are evaluated at the base point $\xz$. Each
holonomically equivalent class of loops will be called a {\it hoop}.
The space of all hoops has, naturally, the structure of a group, which we
call the $G$-{\it hoop group} and denote by $\HG_G$.

It turns out, however, that the structure of $\HG_G$ is largely
insensitive to the specific choice of the Lie group $G$ \cite{AL1}.
More precisely, there are only two hoop groups: An Abelian one for the
case when $G$ is Abelian and a non-Abelian one for the case when $G$
is non-Abelian. In this paper, we will assume that $G$ is {\it
non-Abelian} and denote the corresponding hoop group simply by $\HG$.
(The Abelian case is discussed in some detail in Appendix A of
\cite{AL1}.) $\HG$ can be described purely in terms of
the geometry of the underlying manifold $\S$: Two loops $\a$ and
$\b$ in $\Lx$ define the same element in $\HG$ if they are related
either by a reparametrization or by retracing of a line segment.  More
precisely, we have the following. An unparametrized loop is a class of
loops any two elements of which differ by a piecewise analytic and
orientation preserving diffeomorphism $[0,1]\ \rightarrow\ [0,1]$ and
two unparametrized loops $\a$ and $\b$ define the same hoop in $\HG$,
iff they can be written as
$$\a \ =\ p_2\circ p_1, \ \ \ \b\ =\ p_2\circ q^{-1}\circ
q\circ p_1$$
for some curves $p_i$ and $q$.

Next, we have the notion of {\it independent} hoops. A set of $n$
loops $(\a_1, ...,\a_n )$ in $\Lx$ will be said to be independent if
each $\a_i$ contains an open segment which traversed only once and
which is shared by any other loop at most at a finite number of
points. A set of $n$ hoops is said to be independent if one can find a
representative loop in each hoop such that the resulting collection of
$n$ loops is independent. A subgroup $\g$ of $\HG$ is said to be {\it
tame} if it is generated by a finite number of independent hoops.  It
is straight forward to show that every subgroup of $\HG$ generated by
a finite number of hoops (which are not necessarily independent) is in
fact contained in a tame subgroup. \cite{AL1}.

Tame subgroups $\g$ of $\HG$ will serve as labels for our projective
family. The partial ordering is given just by the inclusion relation.
To show that this set is directed, i.e. satisfies Eq (2.b), we proceed
as follows.  Given any two tame subgroups $\g$ and $\g'$, denote by
$\tilde{\g}$ the subgroup of $\HG$ generated by the independent
generators of $\g$ and $\g'$. Since $\tilde{\g}$ is finitely
generated, it is contained in a tame subgroup $\g''$ of $\HG$.  Hence,
in particular, there exists a tame group $\g''$ such that $\g'' \ge
\g$ and $\g'' \ge \g'$.  Finally, note that since the labels $\g$ are
all {\it finitely} generated, the hoop group $\HG$ does {\it not}
belong to the label set $L$. In particular, therefore, $L$ does not
contain a largest element.

\medskip

\noindent{\bf The projective family}

To each $\g \in L$, we
assign the set $\X_\g$ as follows:
$$\X_{\g}\ :=\ \Hom (\g ,G).\eqno(3.1)$$
The projection maps are the obvious ones: if $\g' \ge \g$, we have
the natural restriction map
$$p_{\g \g'}\:  \Hom(\g',G)\rightarrow \Hom(\g, G)\ .\eqno(3.2)$$
Now, a key property of the tame subgroups of $\HG$ is that a given $\g
\in L$, every homomorphism from $\g$ to $G$ is extendable to the
entire $\HG$ \cite{AL1}. Hence, in particular, it is extendable to
every $\g'\ge \g$.  Therefore, for any $\g, \g' \in L$, the projection
$p_{\g\g'}$ is surjective as required in section 2.1.

Next, we will show that, for each $\g$, the space $\X_{\g}$ inherits
from $G$ the structure of a compact, analytic manifold. Choose a set
$\b_1, ...,\b_n$ of independent generators of $\g$ and consider the
map
$$\X_{\g}\ni H\ \mapsto\ \ (H(\b_1),...,H(\b_n))\in G^n\ .
\eqno(3.3)$$
This map is bijective and provides $\X_\g$ with the manifold structure
of $G^n$. It is easy to verify that this structure is insensitive to
the initial choice of the generators $(\b_1, ...,\b_n)$. Moreover, the
projections $p_{\g'\g}$ are analytic with respect to the induced
manifold structures on $\X_\g$ and $\X_{\g'}$ respectively.

Thus, $(\X_{\g} , p_{\g\g'})_{\g,\g'\in L}$, where $\X_{\g}$ is
regarded as a compact, Hausdorff space, constitutes a compact,
Hausdorff projective family in the sense of section 2.1. We will refer
to it as the $\HG$-projective family.

Finally, note that the structure group $G$ has a natural action on
each component $\X_{\g}$ of the projective family:
$$\X_{\g}\ni H \ \mapsto \ {\rm Ad}(g)\circ H\in \X_{\g},\ \
\forall g\in G \ , \eqno(3.4)$$
where $({\rm Ad}(g)\circ H)(\b) = g^{-1}\cdot H(\b)\cdot g, \ \forall
\ \b\in \g$.  It is easy to check that, in the terminology of section
2.4, $G$ acts consistently on our projective family $(\X_{\g},
p_{\g\g'})_{\g,\g'\in L}$. Therefore, it defines the quotient
$\HG$-projective family. We will denote it by $(\X'_\g, p_{\g
\g'})_{\g,\g'\in L}$, where
$$\X'_\g\ :=\ \X_{\g} /{\rm Ad}(G)\ ,\eqno(3.5)$$
and the projections descend from (3.2) in the obvious way. (As in
section 2.4, we will use the same symbol for the two sets of
projections; the context should suffice to indicate which projection
is intended.)

\medskip\noindent

{\bf The projective limit}

We will first show that the projective limit $\Xb$ of the
$\HG$-projective family can be identified with the set of
homomorphisms of the entire hoop-group $\HG$ into $G$. Pick any
$H\in \Hom(\HG, G)$. For every $\g\in L$, we can restrict $H$ to $\g$
to obtain a homomorphism $H_{|_\s}\in \X_{\g}$ from $\g$ to $G$. It is
obvious that:
$$p_{\g\g'}\ (H_{|_{\g'}})\ =\ H_{|_\g},\ \ \ {\rm whenever}\ \ \
\g'\ge\g.$$
Hence, we have obtained a map
$$\Hom(\HG, G)\ \rightarrow \ \Xb,\ \ \  H\ \mapsto \
(H_{|_\g})_{\g\in L}\ . \eqno(3.6)$$
This is the required identification:

\begin{proposition}{\rm :} The map (3.6) is a bijection.
\end{proposition}

{\it Proof} : That the map is injective as well as surjective follows
easily from the fact that for every hoop $\a\in\HG$ there exists a
tame subgroup $\g\subset\HG$ which contains $\a$ \cite{AL1}.

\qed

We can now extend this result to the quotient $\HG$-projective family.
Since the map (3.6) is $G$-equivariant with respect to the (adjoint)
actions of $G$ on $\Xb$ and $\Hom(\HG,G)$, it projects to a bijective
map of the quotients. Next, consider the map
$$\Hom(\HG,G)/{\rm Ad}(G)\ \rightarrow\ \Xb/G \rightarrow\ \Xpb,
\eqno(3.7a)$$
where $\Xpb$ is the projective limit of the quotient family, given by
$$[H]\ \mapsto\ [(H_{|_\g})_{\g\in L}]\ \mapsto\
([H_{|_\g}])_{\g\in L}. \eqno(3.7b)$$
Theorem 3 now implies:

\begin{proposition}{\rm :} The map (3.7) is a bijection between
$\Hom(\HG, G) /{\rm Ad}(G)$ and the projective limit $\Xpb$ of the
quotient $\HG$-projective family.
\end{proposition}
\bigskip

We conclude this subsection with a remark. A number of results used in
this (as well as the next) subsection depend critically on the
assumption that the loops are all (continuous and) {\it piecewise
analytic}.  (For example, our argument that the tame subgroups of
$\HG$ constitute a directed set fails if the loops are allowed to be
smooth.)  It is not known how much of this analysis would go through
if the loops were assumed to be only piecewise smooth.  If the
structure group is Abelian, piecewise smoothness does suffice (see
Appendix A in \cite{AL1}). However, the methods used there are tied to
the special features of the Abelian case.

\subsection{\C*-algebras of cylindrical functions}

In this sub-section, we will obtain a characterization of the Gel'fand
spectrum of the holonomy \C*-algebra $\HAbar$ generated by the
Wilson loop functions on the space $\ag$ of connections modulo gauge
transformations.

Fix a principal bundle $(P,\S, \pi, G)$ over $\S$ with structure group
$G$. Consider the fiber $\pi^{-1}(x_0)\subset P$ over the base point
$x_0$ used in the hoop group, and fix a point ${\tilde x}_0$ in it.
Then, given a hoop $\a$, each smooth connection $A\in \A$ defines an
element $H(A, \a)$ of the structure group $G$, via holonomy. This
map is in fact a homomorphism of groups. Thus, we have the map
$$\A\ni A\ \mapsto\ H^A\in \Hom(\HG, G),\ \ \ H^A(\a)\ =\
H(A,\a)\eqno(3.8)$$
from the space $\A$ of connections to the space of homomorphisms from
$\HG$ to $G$. Now, given a tame subgroup $\g$ of the hoop group, (3.8)
induces a map from $\ag$ to the space $\Hom (\g , G)/{\rm Ad}G$:
$${\tilde p}_\g\:\ \ag\ \rightarrow\  \Hom(\g,G)/{\rm Ad}(G)\ \ \
{\tilde p}{_\g} ([A])\ := \ [{H^A}_{|_\s}].\eqno(3.9)$$
This map is known to be surjective \cite{AL1}. (It is, in effect, the
projection from the space of classical configurations/histories of the
continuum theory to the corresponding space for the lattice theory
associated with $\g$.)  The pull-back ${{\tilde p}_\s}^*$ of this map
carries functions on $\X'_{\g}$ $(=\Hom(\g,G)/ {\rm Ad}(G))$ to
functions on $\ag$. We can use these pull-backs to introduce the
notion of cylindrical functions on $\ag$. Thus, the space $\Cyl(\ag)$
of {\it cylindrical functions} on $\ag$ is defined to be
\cite{AL1}
$$\Cyl(\ag)\:=\ \bigcup_{\g\in L}\ {{\tilde p}_\s}^*\ C^0(\X'_{\g}).
\eqno(3.10)$$
Since each $\X'_{\g}$ is in particular compact and Hausdorff,
$\Cyl(\ag)$ has the structure of a normed $\star$-algebra. Its
completion, $\Cylb(\ag)$, will be called the {\it \C*-algebra of
cylindrical functions} on $\ag$.

A basic example of a cylindrical function on $\ag$ is the Wilson loop
function. Fix a representation $\rho:G\rightarrow {\rm End}(V)$ of $G$
where $V$ is an $n$-dimensional vector space, and use it to define
traces. Then, a Wilson loop function $W_\a$ on $\ag$, labelled by a
hoop $\a$, is defined by $W_\a ([A]) :=\textstyle{1\over n} {\rm Tr}
\rho (H(\a, A))$. To see that this is a cylindrical function, choose
any $\g\in L$ which contains $\a$. Then, $W_\a([A])$ is the pull-back
of the function $f_{[A]}$ on $\X'_{\g}$ defined by $f_{[A]}([H]):=
\textstyle{1\over n} \Tr H$; it is a cylindrical function on $\g$.
Indeed, as one might intuitively expect, Wilson loop functions
generate the entire algebra of cylindrical functions: $\Cylb(\ag)$ is
naturally isomorphic to the holonomy \C*-algebra $\HAbar$ generated by
the Wilson loop functions on $\ag$. (For the case when $G$ is $SU(n)$
or $U(n)$, see \cite{AL1}. For the general compact, connected gauge
group, only a small modification required; one has to allow, all the
fundamental representations.)  The problem of characterizing the
Gel'fand spectrum of the holonomy algebra therefore reduces to that of
characterizing the spectrum of the \C*-algebra $\Cylb(\ag)$.
The task is further simplified by the following result:

\begin{proposition}{\rm :} The \C*-algebra $\Cylb(\ag)$ of cylindrical
functions on $\ag$ is isometrically isomorphic with the \C*-algebra
$\Cyl(\Xpb)$ of cylindrical functions on the quotient $\HG$-projective
family. The isomorphism is the continuous extension of the map
$$\Cyl(\ag)\ni {{\tilde p}_\s}^*\ f_\s\ \mapsto\ [f_\s]_\sim\in
\c(\Xpb)\ . \eqno( 3.11)$$
\end{proposition}

{\it Proof} : The result follows easily from the fact that the maps
${\tilde p}_\s$ of (3.9) are surjections.

\qed
\medskip

Using Theorem 1, we now have the desired characterization of the
Gel'fand spectrum of $\HAbar$:

\begin{theorem}{\rm :} The Gel'fand spectrum of the \C*-algebra
$\Cylb(\ag)$ is naturally homeomorphic with the space $\Xpb\ =\
\Hom(\HG,G)/Ad(G)$.  The homeomorphism assigns to $[H]\in\Xpb$ the
following functional defined on $\Cylb(\ag)$:
$$ [H](f)\ :=\ f_\s([H]_{|_\s})\eqno(3.12)$$
for any $f_\s\in C^0(\X'_\g)$ such that $f\ =\ {{\tilde p}_\s}^*\ f_\s$.
\end{theorem}

Let us summarize. Proposition 5 implies that $\Cylb(\ag)$ --and hence
the holonomy \C*-algebra $\HAbar$-- is just a faithful representation
of the \C*-algebra $\Cylb(\Xpb)$ of cylindrical functions associated
with the quotient $\HG$-projective family. Theorem 1 ensures us that
the Gel'fand spectrum $\agb$ of $\HAbar$ can be identified with the
projective limit $\Xpb$ of the quotient $\HG$-family. Now, since
elements of $\HAbar$ suffice to seperate points of $\ag$, and since
$\agb$ is the spectrum of $\HAbar$, basic results of the Gel'fand
theory imply that $\ag$ is densely embedded in $\agb$.  Hence, it now
follows that $\ag$ is densely embedded in $\Xpb \equiv \Hom(\HG, G)/
{\rm Ad}(G)$.  The projections ${\tilde p}_{\g} : \ag \rightarrow
\X'_\g$ of (3.9) are just the restriction to $\ag\subset\Xpb$ of the
projections ${ p}_{\s} : \Xpb\rightarrow \X'_\g$. Hence the space of
the cylindrical functions on $\ag$ may be viewed as the restriction to
$\ag\subset \Xpb$ of the cylindrical functions on $\Xpb$, defined in
terms of the projective family. Finally, we have an independent proof
of the Marolf and Mour\~ao \cite{MM} result that the Gel'fand topology
on $\agb$ coincides with the Tychonov topology on $\Xpb$, a proof that
now holds for general compact, connected gauge groups.

Recall that, given a tame subgroup $\g$ of $\HG$ generated freely by
$n$ independent hoops, the member $\X'_{\g}$ of the projective family
is isomorphic with $G^n/{\rm Ad}(G)$. Hence, we can think of it as the
space of configurations (or histories) of the gauge theory associated
with the lattice represented by $\g$. The projection maps
$\tilde{p}_S$ just serve to reduce the continuum theory to this
lattice theory; they ignore what the connections do at points which
are not on the lattice represented by $\g$. As we enlarge the subgroup
$\g$, the lattice theory captures more and more information contained
in the connection. The full information is recovered in the projective
limit. Thus, this approach to the continuum theory is tied closely to
the lattice approach. That, in particular, is the underlying reason
why we can maintain manifest gauge invariance.  However, the continuum
limit is taken in a somewhat non-standard way: rather than refining
the ``mesh'' of a rectangular lattice further and further, one allows
bigger and bigger tame subgroups of the hoop group.

\subsection{Measures on $\agb$}

We are now ready to introduce measures on $\agb$. The basic idea is to
apply results of section 2.3 to the $\HG$-projective family
constructed above (using techniques from \cite{Ba1,MM}). Our aim in
this subsection is to present only the key elements of the actual
constructions of various measures and to provide the overall picture;
a detailed account would make this article inordinately long.

Of the measures we will introduce, there are several families that are
invariant under the induced action of the diffeomorphism group of
$\S$.  It is therefore convenient to first spell out what this
property entails. Consider an analytic diffeomorphism $\Phi$:
$$\Phi\:\ \S\ \rightarrow\ \S\ .\eqno(3.13a)$$
Since $\Phi$ has a well-defined action on the space of continuous,
piecewise analytic loops, and since the action obviously preserves the
hoop equivalence relation, it induces an isomorphism on $\HG$ which in
turn induces an isomorphism on the quotient $\HG$-projective family:
$$\Phi^\g\:\ \X'_{\Phi(\g)}\ \rightarrow\ \ \X'_{\g}\ .\eqno(3.13b)$$
Each $\Phi^\g$ is a homeomorphism and this family of homeomorphisms
induces a homeomorphism of the projective limit $\Xpb$ on to itself.
The question then is if a given measure $\mu$ on $\Xpb$ is
diffeomorphism invariant. It is easy to check that it is so if and only
if the corresponding family $\mu_{\g}$ of measures on $\X'_{\g}$
satisfies \cite{Ba1}:
$$ \Phi^{\g}_* \  \mu_{\Phi(\g)}\ = \ \mu_{\g} \ .\eqno(3.13c)$$

\medskip

\noindent{\bf The induced Haar measure}

Let us begin with the $\HG$-projective family $(\X_{\g},
p_{\g\g'})_{\g, \g' \in L}$.  Recall from section 3.1 that, if the
tame subgroup $\g$ of $\HG$ is generated by $n$ independent hoops,
then $\X_{\g}$ is homeomorphic with $G^n$. Given a specific choice of
$n$ independent generators of $\g$, the explicit homeomorphism is
given by Eq (3.3).  Denote by $\mu_H$, the normalized Haar measure on
$G$.  The homeomorphism pushes the induced Haar measure on $G^n$
forward and provides us with a measure on $\X_{\g}$. Using properties
of the Haar measure, it is easy to verify that this measure on
$\X_{\g}$ is insensitive to the initial choice of the independent
generators. We will denote it by $\mu_H^{(n)}$.  Thus, each member of
our $\HG$-projective family is equipped with a regular, Borel,
probability measure. It is easy to verify that this family
$\mu_H^{(n)}$ of measures is consistent in the sense of Eq (2.21).
Hence, by Theorem 2, it defines a regular, Borel, probability measure
$\mu_o$ on the projective limit $\Xb$.  Since the Haar measure is
faithful, so is the measure $\mu_H^{(n)}$ on $\X_{\g}$ for any tame
subgroups $\g$. Hence, again by Theorem 2, it follows that $\mu_o$ is
also faithful.

Next, consider the quotient $\HG$-family. Since its projective limit
$\Xpb$ can be expressed as $\Xpb = \Xb/G$, we can push forward $\mu_o$
on $\Xb$ to obtain a measure $\mu_o'$ on $\Xpb \equiv \agb$. This is
the required induced Haar measure. It is a faithful, regular, Borel,
probability measure on $\agb$. Finally, since its construction did not
involve any structure --such as a metric, a fiducial connection, or a
volume element on $\S$-- it follows that $\mu'_o$ is invariant under
the induced action of Diff-$\S$.

This measure was introduced in \cite{AL1}, where further details can
be found.

\medskip\goodbreak
\noindent{\bf Measures generated by knot invariants}

One can use the induced Haar measure $\mu'_o$ to obtain a large class
of diffeomorphism invariant measures, one corresponding to each knot
invariant (of regular, embedded loops) of a suitable type. Here, we
will present this construction for the simplest non-trivial case, that
with the structure group $G = SU(2)$.

In this case, any regular, Borel, probability measure on $\agb$ is
completely determined by the generating function of Eq (1.1) involving
only {\it single} loops \cite{AI}:
$$\Gamma_{\mu}(\a) := \int_{\agb}\ W_\a(\bar{A})\ d\mu \ .
\eqno(3.14)$$
Let $k_o$ be the characteristic function of an arbitrarily chosen knot
class, say $\{\b\}_o$ of smoothly embedded loops in $\S$ (with no self
overlaps). Thus, $k_o(\b) = 1$ if $\b$ belongs to $\{\b\}_o$ of regular
loops and zero otherwise. Consider the formal sum $\sum_{\b} k_o(\b)
W_\b (\bar{A})$, over all the loops in $\S$, and set:
$$\Gamma_{\mu(k_o)}(\a) = \int_{\agb}\ \ W_\a (\bar{A})\ [\sum_{\b}
k_o(\b) W_{\b} (\bar{A})]d\mu'_o \ . \eqno(3.15)$$
Now, $SU(2)$ identities imply that $W_{\a} W_{\b}= \textstyle{1\over
2}(W_{\a\circ\b} + W_{\a\circ{\b^{-1}}})$, whence the integrand
reduces to a formal sum of ($k_o$ times) certain Wilson loop
functions. Now, our measure $\mu'_o$ is such that the integral of all
but a finite number of these terms is identically zero. Hence, the
integral can actually be given a rigorous meaning. The resulting
$\Gamma_{\mu(k_o)}$ can be shown to have the properties required to
qualify as the Fourier transform of a signed (i.e. not necessarily
positive definite) measure $\mu_o(k)$ on $\agb$. Thus, while the sum
$\sum_{\b} k(\b) W_\b (\bar{A})$ is only formal and does not define a
function on $\agb$, there is a precise sense in which $\sum_{\b} k_o(\b)
W_\b (\bar{A})\ \times \mu'_o(\bar{A})$ can be regarded as a measure
on $\agb$.

Again, by construction, these measures are all invariant under the
induced action of the Diff$(\S)$ group.  Similar measures can be
constructed if $k_o$ is replaced by a more general knot invariant of
an appropriate type. Details will appear in \cite{ALMMT2}.

\medskip

\noindent{\bf Baez vertex measures}

We will now discuss another family of Diff$(\S)$ invariant measures
introduced by Baez \cite{Ba1,Ba2}, using a projective family labelled
by graphs.

In terms of our projective families, Baez's construction can be
summarized as follows. Given any tame subgroup $\g$ of the hoop group
$\HG$, let us cut the loops in $\g$ to obtain a finite set of
independent, analytic edges (with no overlaps) whose composition
generates all the loops contained in $\g$. (See \cite{AL1} for
details). These edges form a graph which is embedded in $\S$ and which
contains all the loops in $\g$. (Thus, $\g$ is contained in the
fundamental group of the graph.)  To an edge $e$, associate two
$G$-valued random variables, $h_{e-}, h_{e+}$, each of which is
assigned to an endpoint of the edge ---i.e., to a vertex. For each
vertex $v$ in the graph, denote by $n_v$ the number of edges incident
at the point (any edge with both points at $v$ being counted twice)
and consider the associated $n_v$ random variables. The space spanned
by these variables is homeomorphic to $G^{n_v}$. The key idea is to
introduce on $G^{n_v}$ a measure $\mu_v$ which depends {\it only on
the diffeomorphism invariant characteristics of the vertex} $v$.
(Examples of such characteristics are: a cusp at $v$, a corner at $v$,
a simple intersection of two smooth line segments, a multiple
intersection, etc.) Now, given a function $f$ on $\X_{\g}$, we
acquire, via Eq (3.3), a function $f(g_1,..., g_n)$ on $G^n$. By
expressing each independent generator of $\g$ as a composition of
edges, we can reexpress $f$ as a function $f(g_1(h), ..., g_n(h))$ on
the space $G^{n_{v_1}+ ... +n_{v_V}}$ of random variables $h$
associated with all the edges in the graph. Now, we can state the Baez
ans\"atz:
$$\int_{\X_{\g}} f_\g\ d\mu^B_\g\ := \ \int_{G^{n_{v_1}}\times ...
\times G^{n_{v_V}}}\  f(g_1(h),...,g_n(h))\ d\mu_{v_1}(h_{v_1})...
d\mu_{v_V}(h_{v_V})\ ,\eqno(3.16)$$
where $V$ is the total number of vertices in the graph, and, as
before, $h\in G^{n_{v_1}+...+n_{v_V}}$ and $h_v\in G^{n_v}$.  This
ans\"atz simplifies the task of solving the consistency conditions on
the family $\mu_S$ of measures considerably. In particular, one can
construct a solution of these conditions for {\it each} choice of a
probability measure on the structure group $G$. Each solution in this
large class defines a Diff$(\S)$ invariant measure on the projective
limit $\Xpb \equiv
\agb$.

By construction, the Baez maeasures are sensitive only to the presence
of non-trivial vertices in graphs. Consequently, their support is
contained within a proper subset of $(\agb)_B $ of $\agb$:
$$ (\agb)_B := \Hom_B(\HG, G)\ ,\eqno(3.17) $$
where $\Hom_B(\HG, G)$ is the set of all homomorphisms $H$ such that
$H(\a)\ =\ I_G$, whenever a hoop $\a\in\HG$ can be represented by an
analytic embedding of a circle. Thus, the Baez measures are not
faithful. One can of course obtain faithful measures by taking convex
linear combination of any Baez measure with the induced Haar measure
$\mu'_o$.

\medskip

\goodbreak
\noindent{\bf The homotopy measure}

We now turn to a measure \cite{AI} that is of interest to theories
such as 3-dimensional gravity where only the flat connections are of
physical interest. From the viewpoint of the general theory, however,
this measure is rather trivial since it has support on a {\it finite}
dimensional subspace of $\ag$.

Consider the space $\Hom(\pi_1(\S), G)$ of homomorphisms from the
first homotopy group of the manifold $\S$ into the structure group $G$
and equip it with a topology such that for every tame subgroup of
hoops $\g$, the natural map
$$\Hom(\pi_1(\S), G)\ \rightarrow\ \Hom(\g, G)\eqno(3.18)$$
is continuous. Then, for every measure $\mu_f$ defined on
$\Hom(\pi_1(\S), G)$, the push forward to $\Hom(\HG, G)$ defines a
measure on $\Xpb$, which we will denote by $\mu_F$. The space
$\Hom(\pi_1(\S),G)/Ad(G)$ corresponds, of course, to the space of flat
connections over $\S$.  Hence, the interpretation of the action of the
measures $\mu_F$ on functions on $\agb$ is as follows: given a
function on $\agb$, one first restricts it to a function on the space
of flat connections and integrates this restriction using $\mu_f$.

The measure $\mu_F$ is invariant with respect to group ${\rm
Diff}_o(\S)$ consisting of diffeomorphisms of $\S$ which are generated
by analytic vector fields. Finally, we note that one can generalize
this example by replacing the flat connections by those which are
reducible (up to the conjugacy) to a given subgroup of $G$.

\medskip

\noindent{\bf Heat-kernel measures.}

The measures discussed above are all invariant under Diff$(\S)$ and
therefore of interest primarily to diffeomorphism invariant theories
such as general relativity. (As discussed below, however, $\mu'_o$ can
also serve as a fiducial measure in the continuum Yang-Mills theory.)
We will now discuss a class of measures \cite{AL2} which do not share
this invariance and may be useful in physical theories which, e.g.,
depend on a background space-time metric. On the mathematical side,
these measures are, in a certain sense, the natural non-linear analogs
of the Gaussian measures on linear spaces.

We begin by noting that since the structure group $G$ is a compact,
connected Lie group, it admits a family of heat kernel measures
$\mu_t$, with $t>0$. These are obtained as follows. One first solves
the heat equation on $G$:
$$ {d \rho_t\over dt} = {1\over 2}\ \bigtriangleup \rho_t\ , \quad
\rho_{t=0}(g) = \delta(g, 1_G) \eqno(3.19)$$
where $\triangle$ is the natural Laplacian on $G$ and $1_G$ is the
identity in $G$. It is known that $\rho_t$ is strictly positive and
smooth for all $t>0$ \cite{S}. (An explicit expression for $\rho_t$ in
terms of the characters of $G$ is also available.) Hence, we can set
$\mu_t = \rho_t \mu_H$. These are the heat-kernel measures on $G$.

The idea is to use these to obtain measures on $\agb$. It turns out,
however, that to satisfy the consistency conditions, it is necessary to
introduce additional structure on the space of analytic curves, namely
a ``length functional''. More precisely, consider a positive function
$l$ on the space of finite, unparametrized, anaytic curves on $\S$,
satisfying:
$$l(e) = l(e^{-1}),\ \ \ l(e_1\circ e_2) = l(e_1) + l(e_2),
\eqno(3.20)$$
for all curves $e, e_1, e_2$. If $\S$ is equipped with a positive
definite metric, one can let $l(e)$ be just the length of the edge
with respect to this metric. There are, however, many more solutions
to (3.20).

Given a ``length functional'' we can introduce a measure on the
members $\X_{\g}$ of the $\HG$-projective family as follows. Fix a
tame subgroup $\g$ of the hoop group. As in the case of the Baez
measures, introduce a graph which contains the group $\g$ as a
subgroup of its fundamental group. To each edge $e_i$ in the graph
assign a $G$-valued random variable $h_i$ and, for every $t >0$, a
heat kernel measure $\mu_{h_i} = \rho_{s_i} \mu_H(h_i)$, with $s_i=
l(e_i)t$, on $G$. Then, using the notation of Eq (3.16), we can define
a measure $\mu_\g$ on $\X'_\g$ as follows:
$$\int_{\X'_{\g}}\  f\  d\mu_\g\ := \ \int_{G^E}\
f(g_1(h),...,g_n(h))\ \mu_H(h_1)....\mu_H(h_E) \ ,\eqno(3.20)$$
where $f_\g\in C^0({\X'}_\g)$, and where $E$ denotes the number of
edges in the graph. This family of measures is well-defined and
consistent.  Hence, it defines a regular, Borel, probability measure
on $\agb$. Note, however, that there is no non-trivial ``length
functional'' which is Diff$(\S)$ invariant.  Hence none of these
heat-kernel measures on $\agb$ are Diff$(\S)$ invariant.

Remarkably, however, with each of these heat-kernel measures, there is
an associated Laplace operator defined on $\agb$ which is essentially
self adjoint on $L^2(\agb, \mu'_0)$ \cite{AL2}. Since such operators
generally feature in the Hamiltonians of field theories, these
measures may well play an important role in the canonical quantization
of such theories.  Finally, the heat-kernel measures can be used to
construct the analog of the Segal-Bargmann transform of linear quantum
field theories which provides a {\it holomorphic} representation of
quantum states
\cite{ALMMT3}.

\medskip

\noindent{\bf Measures for 2-dimensional Yang-Mills theories}

Finally, let us discuss Yang-Mills theories. The $SU(n)$ Yang-Mills
theories have been successfully quantized in two dimensions using
measures on $\agb$, in the cases when the topology is $R^2$ or
$S^1\times R$ \cite{ALMMT1} (see also the earlier work of Klimek \&
Kondracki \cite{KK}).  Here we will sketch the construction of the
Euclidean theory.

The idea, as in constructive quantum field theory, is to begin with a
fiducial measure, but now on $\agb$. This is chosen to be the natural
measure $\mu'_o$. Heuristically, the correct, physical measure may be
written as $\mu_{\rm phy} = \exp\ -I({\bar A}) \mu'_o$. However, since
the Yang-Mills action $I$ is not well-defined on generalized
connection, one must first introduce a regularized version thereof.
The Wilson lattice action is an especially natural candidate for this
since it is an integrable function on $\agb$. One therefore introduces
a lattice with lattice-separation $a$, writes down the associated
Wilson action $I_a$, and computes the integrals
$$\Gamma_{(a)}(\a_1,...,\a_n) = \int_{\agb}\
W_{\a_1}(\bar{A})...W_{\a_n}(\bar{A})\ \exp (-I_a(\bar{A})) \ d\mu'_o
\ .\eqno(3.21)$$
This integral is well-defined; in fact the integrand is a cylindrical
function on $\agb$. One now takes the ultraviolet limit (lattice
separation goes to zero) as well as the thermodynamic limit (lattice
covers the entire Euclidean space-time) of $\Gamma_{(a)}$. The result
is a well-defined function of multi-loops which satisfies all the
conditions to qualify as the Fourier transform of a regular, Borel
probability measure on $\agb$.  (In the case of the $U(1)$-theory, the
Fourier transform can be written out in a closed form.) This is the
physical measure of the theory.

This method interacts especially well with lattice approximations. In
particular, it provides a natural framework for the continuum theory
which one hopes will emerge from lattice methods in higher space-time
dimensions. We should empahsize, however, that since higher
dimensional Yang-Mills theories have not yet been analyzed in this
framework, it is not yet clear if our general strategy will continue
to be viable there.

\medskip

\section{Discussion}

In this paper, we used projective techniques to construct a manifestly
gauge invaraint framework for functional integration in gauge
theories. Most of the final results reported in this paper have
already appeared in the literature during the last two years.
However, the approach adopted is new and it has enabled us to provide
an essentially self-contained, concise and significantly simpler
treatment of this material.

A key strength of these methods lies in the fact that they face the
``kinematic non-linearities'' of gauge theories squarely. More
precisely, one recognizes early on that the space $\ag$ of {\it
physically distinct} configurations or histories in gauge theories
fails to admit a vector space structure and then takes these
non-linearities seriously. This is to be contrasted with the more
familiar approaches that attempt to fix a suitable gauge --ignoring
Gribov ambiguities-- to force a linear structure on $\ag$, and then
look for measures which are perturbations of a (free) Gaussian measure
on a linear space. The measures we discussed, by contrast, are all
rather ``far'' from the ones obtained in such perturbative treatments.
They are genuinely non-perturbative, geared to the kinematical
non-linearities of gauge theories. As pointed out in section 3, the
approach is closely related to the lattice approach; {\it the
kinematics of the continuum theory is recovered from the lattice
theories in a projective limit.}

Although these methods themselves are thus quite different from the
ones used in the linear theories, as we saw in this paper, they do
fall in the broad category of projective techniques. Thus, at a
``primary'' level, there {\it is} a structural similarity between our
approach and the one based on promeasures \cite{DM1,DM2,DM3}, used in
the linear case. In both cases, the basic objects are the projective
families and one is ultimately interested in projective limits. The
difference lies in the specific projective families used. The members
of the family used in the linear cases are all finite dimensional
vector spaces. In our case, they are compact, Hausdorff topological
spaces. This difference does have important consequences; indeed even
the sets of natural questions in the two cases are different.
Nonetheless, when looked at from a sufficiently abstract perspective,
one does find an underlying unity and coherence. Our non-linear
generalization is thus not arbitrary; there is a precise sense in
which it is a natural extension of the classical techniques \cite{K}
for functional integration on linear topological spaces.

\bigskip
{\bf Acknowledgements}: We would like to thank John Baez, Donald
Marolf, Jose Mourao and Thomas Thiemann for discussions. Jerzy
Lewandowski is grateful to Center for Gravitational Physics at Penn
State and Erwin Schr\"odinger Institute for Mathematical Physics in
Vienna, where most of this work was completed, for warm hospitality.
This work was supported in part by the NSF grants 93-96246 and
PHY91-07007, the Eberly Research Fund of Penn State University, the
Isaac Newton Institute, the Erwin Shr\"odinger Institute and by the
KBN grant 2-P302 11207.

\end{document}